\documentclass[aps,prl,reprint,groupedaddress,superscriptaddress,floatfix,10pt]{revtex4-1}

\usepackage[english]{babel} 
\usepackage[utf8]{inputenc}
\usepackage[T1]{fontenc}
\usepackage{lmodern}

\usepackage{amsmath}
\usepackage{amssymb}
\usepackage{mathptmx}
\usepackage{graphicx}
\usepackage{float}
\usepackage{caption}     
\usepackage{color}
\usepackage{bm}
\usepackage{braket}

\usepackage{hyperref}

\usepackage{color}
\usepackage{ulem}

\newcommand{\I}{\mathrm{i}}
\newcommand{\E}{\mathrm{e}}

\newcommand{\Ec}{E_{\mathrm{c}}}		
\newcommand{\Eg}{E_{\mathrm{g}}}
\newcommand{\Kc}{K_{\mathrm{c}}}
\newcommand{\nE}{n_{\mathrm{e}}}	
\newcommand{\QL}{\hat{Q}_{\mathrm{L}}}
\newcommand{\wL}{\omega_{\mathrm{L}}}
\newcommand{\wB}{\omega_{\mathrm{B}}}
\newcommand{\rmL}{\mathrm{L}}

\begin{document}
\captionsetup[figure]{labelfont={},labelformat={default},name={Fig.}}

\title{Nonclassical light generation and control from laser-driven semiconductor intraband excitations}

\author{Ivan Gonoskov}
\email[]{ivan.gonoskov@uni-jena.de}
\thanks{These authors contributed equally}
\affiliation{Institute of Physical Chemistry, Friedrich Schiller University Jena, Helmholtzweg 4, 07743 Jena, Germany}

\author{Ren\'e Sondenheimer}
\email{rene.sondenheimer@iof.fraunhofer.de}
\thanks{These authors contributed equally}
\affiliation{Fraunhofer Institute for Applied Optics and Precision Engineering, Albert-Einstein-Strasse 7, 07745 Jena, Germany}

\author{Christian H\"unecke}
\affiliation{Institute of Physical Chemistry, Friedrich Schiller University Jena, Helmholtzweg 4, 07743 Jena, Germany}

\author{Daniil Kartashov}
\affiliation{Institute of Optics and Quantum Electronics, Friedrich Schiller University Jena, Max-Wien-Platz 1, 07743 Jena, Germany}

\author{Ulf Peschel}
\affiliation{Institute of Solid State Theory and Optics, Friedrich Schiller University Jena, Max-Wien-Platz 1, 07743 Jena, Germany}

\author{Stefanie Gräfe}
\email{s.graefe@uni-jena.de}
\affiliation{Institute of Physical Chemistry, Friedrich Schiller University Jena, Helmholtzweg 4, 07743 Jena, Germany}
\affiliation{Fraunhofer Institute for Applied Optics and Precision Engineering, Albert-Einstein-Strasse 7, 07745 Jena, Germany}

\date{\today}

\begin{abstract}
We investigate the generation of higher-order harmonics from a quantum optics perspective via the interaction of a semiconductor with a coherent pump field focusing on the regime where strong-field intraband excitations dominate. The related Schr\"odinger equation for the system is solved approximately and the classical and quantum characteristics of the fundamental light mode as well as the higher frequency modes are analyzed. We find intricate but sufficiently mild modifications of the fundamental mode and coherent displacements depending on the position quadrature component of the driving laser field for the harmonic modes within our approximations due to the intricate induced nonlinear interactions. Similar to high-harmonic  generation in atoms, all radiation field modes are entangled, allowing for potential novel protocols for quantum information processing with high photon numbers over a large range of frequencies.
\end{abstract}


\maketitle

\section{Introduction}
The interaction of matter with light is typically described semiclassically: while the matter is studied within a quantum description, the light is treated classically as electromagnetic fields/waves without considering the contribution of individual photons. This is even more true for the case of intense light interacting with matter \cite{s0,s1,s2}.
The prevalent reason for such an ansatz is based on the classical field limit declaring that for intense fields, quantum corrections are negligible compared to the classical averages \cite{Mandel}. 
Recently, this paradigm began to shake due to pioneering experiments demonstrating quantum properties of the emitted radiation in the highly nonlinear process of high-order harmonic generation (HHG) in gases and bulk \cite{Paris1, Paris2}. This raises the fundamental research question to which extent quantum features prevail in the nonlinear frequency generation processes involving many photons of different frequencies. Moreover, it paves the way towards promising applications such as new measurement techniques \cite{Paris1,Paris2}, laser-harmonic and harmonic-harmonic entanglement \cite{ParisBN}, and the generation of resource states for quantum information processing \cite{ParisN}. However, the description of a complex quantum system which includes one or more extra degrees of freedom (one for each quantized-light mode) faces theoretical challenges and has been only solved for a few special cases \cite{ScR,Mak,T,Fold}. Recently, HHG in atoms was considered from a quantum information theory perspective based on positive operator valued measures \cite{Stammer}. This allows for a deeper understanding of potential novel state engineering via HHG \cite{Lewenstein, Stammer2}.

An advantage for the analysis of the quantum properties of light produced in HHG in atoms is given by the fact that approximative solutions can be found for the quantum optical states \cite{ParisBN,ScR}. For instance, this can be traced back to the fact that the field operator enters the effective interaction Hamiltonian in a linear manner via the dipole approximation \cite{Rivera-Dean:2021rnp}. 

In this letter, we investigate the corresponding quantum states generated in HHG via the interaction of coherent light with a semiconductor. In contrast to the atomic case, for semiconductors, the field operators enter the Hamiltonian in a nonlinear manner according to the band dispersion. We construct a particular - but in the context of semiconductor HHG typical - solution describing the intense field-semiconductor interaction at the quantum level. Hereby, we focus on the lower-order harmonics with energies of quanta below the band gap of the semiconductor. These are, mechanistically, described to arise within the to so-called intraband currents: upon strong-field interaction, electrons are excited from the valence band to the conduction band(s) and are there driven by the intense laser field. Due to the nonlinear band dispersion, this motion leads to the laser-induced intraband current giving rise to pronounced high-order harmonics with photon energies below the band gap. Higher-harmonics, with energies above the band gap, are mechanistically described to be caused by inter-band contributions due to the nonlinear polarization between driven electrons in the conduction and corresponding holes in the valence band. In this work, we focus on the intraband currents, thus (lower) harmonics below the band gap. For a review and description of HHG in semiconductors from a classical perspective, e.g., see \cite{Koch}. For comparison, we analyze the properties of the generated radiation fields at the classical and quantum level. In general, we find non-Gaussian modifications of the radiation field governed by the nonlinear interactions. However, these modifications have a rather peculiar structure such that expectation values of any function of the position quadrature of the pump field are not altered. 

The paper is organized as follows: We first introduced the theoretical description of HHG in semiconductors from a quantum perspective and discuss the related approximations and their limitations. Afterwards, we introduce our analytical method and present the approximate analytical solution of the problem. We compare our analytical solution with a numerical solution obtained from the semiconductor-Maxwell-Bloch equations (SMBE) in the classical regime. Finally, we investigate the non-classical properties of the radiation field produced in HHG under the conditions of the laser-induced intraband excitations.

\section{Theoretical Description}
In order to describe the interaction of a semiconductor with a multi-mode quantized electromagnetic field, we first assume a number of electrons $n_{\mathrm{e}}$ being excited solely into the lowest conduction band. We further assume that these are driven solely in this conduction band, without transitions to other bands. In other words, the process can be described by the band dispersion of the conduction band $\Ec(\vec{K})$ interacting with a quantized radiation field without interband transitions. This is the common scenario, e.g., when a prepulse has excited electrons into the conduction band. 
Alternatively, the driving pulse itself produces mostly intraband excitations.
Moreover, we assume that all the considered intraband electrons move synchronously in the strong field of the driving laser, forming the quasi-classical intraband current (similar as in the classical-electron approximation, see, e.g. \cite{Koch}). In this case, we can approximate the system by a single particle Hamiltonian with respect to the Bloch electrons of the system using the conduction band dispersion $\Ec(\vec{K})$:
\begin{equation}
\hat{H}  =  \nE\; \Ec\!\left(\hat{\vec{p}}-\sum_{j}\frac{e}{c}\hat{\vec{\mathcal{A}}}_{j}\right) + \sum\limits_{j}\omega_{j}\hat{N}_{j},
\label{eq:Hamiltonian2}
\end{equation}
where $\hat{\vec{p}}-\sum_{j}\frac{e}{c}\hat{\vec{\mathcal{A}}}_{j}$ is the canonical momentum of a single electron in the band with $\hat{\vec{\mathcal{A}}}_{j}$ the vector potential operator of the field mode $j$ (we are using units in which $\hbar =1$ and $m_{\mathrm{e}}=1$). Further, $\hat{\vec{p}}$ and $e$ are the momentum and charge of the identical electrons. The interaction-free part of the quantized electromagnetic field modes $j$ is encoded in the terms $\omega_{j}\hat{N}_{j}$, with $\omega_{j}$ the corresponging frequency and $\hat{N}_{j}$ the photon number operator where we neglected an unimportant constant vacuum energy term. 
Note that we assume that all electrons with fixed, time-independent density $\nE(t) = \nE = \mathrm{const}$ have the same momentum. This approximation is justified for an optical excitation as electrons are generated at one instance within the half cycle and with zero momentum at the $\Gamma$ point. 
Second, we employ the dipole approximation (no coordinate dependence in the description of the laser field and its harmonics), and neglect propagation effects (considering a relatively thin crystal). Finally, we assume that the depletion of the driving laser field is small. The considered assumptions allow us to use the so-called parametrical-connection approximation, a method described and developed in \cite{T}. The approximate solution obtained from this method describes the quantum evolution of the light field by treating the intraband-current back-reaction as a finite-order perturbation.

\section{solution method}
In order to obtain an analytical solution of the problem within the approximations described above, it is convenient to transform into the interaction picture with respect to the radiation modes by applying the linear unitary transformation $\prod_{j} \mathrm{exp}(-\mathrm{i}\,t\,\omega_{j}\hat{N}_{j})$ \cite{Mandel,T}. The interaction Hamiltonian then reads,
\begin{align}
\hat{H} = \nE\; \Ec\!\bigg(\hat{\vec{p}}-\sum_{j}\frac{e}{c}\hat{\vec{A}}_{j}(t)\bigg),
\label{eq:Hamiltonian3}
\end{align}
and $\hat{\vec{A}}_{j}(t) = \mathrm{exp}(\I\,t\,\omega_{j}\hat{N}_{j})\; \hat{\vec{\mathcal{A}}}_{j}\; \mathrm{exp}(-\I\,t\,\omega_{j}\hat{N}_{j})$ is the transformed field operator.
Now, we can apply the idea of the parametrical-connection approximation (see \cite{T} for full details). In fact, the effect of the laser-induced current in the conduction band back on to the initial light field is small, i.e., the canonical momentum in the relevant reference frame is mainly determined by the driving radiation field. Thus, the momentum $\hat{\vec{p}}$ of the electrons is negligible such that we are able to solve the reduced time-dependent Schr\"odinger equation (TDSE) (electron states will be averaged out) for the light states separately. Separating the full system simplifies the calculations and subsequent analysis substantially. After applying the parametrical-connection approximation, we consider the evolution of the quantized light modes. We denote the wave function of the entire radiation field with $\ket{G}$ encoding all modes distinguished by the $j$-indices. The TDSE for the non-affected vacuum propagation in the interaction picture is $\mathrm{i}\partial_{t}\ket{G} = 0$. The TDSE which includes the first-order corrections due to quantum back-action of the intraband current reads (see \cite{T}):
\begin{equation}\label{eq:TDSE}
\I \frac{\partial}{\partial t} \ket{G} = \nE\; \Ec\!\bigg( \!\! -\sum_{j}\frac{e}{c}\hat{\vec{A}}_{j}(t)\bigg) \ket{G}.
\end{equation}

In what follows, we will use a coordinate/position quadrature representation for the description of the field vector potential operators as well as for the final state after the interaction of the pump laser with the semiconductor \cite{Mandel,T}, as it will be convenient for the analysis. 
We denote the frequency of the driving field by $\wL$ and the frequencies of the generated higher harmonics by $\omega_{j} = j \wL$. The position and momentum quadratures of the different field modes are given by $\hat{Q}_{j} = \frac{1}{\sqrt{2}}(\hat{a}_{j} + \hat{a}_{j}^{\dagger})$ and $\hat{P}_{j} = \frac{1}{\sqrt{2}\I}(\hat{a}_{j} - \hat{a}_{j}^{\dagger})$, respectively. Note that we will highlight the impact of the fundamental pump mode ($j=1$) by a subscript $\mathrm{L}$, e.g., $\hat{Q}_{1} \equiv \QL$. 
In the position quadrature representation, we obtain $\hat{a}_{j}=\frac{1}{\sqrt{2}}\Big(Q_{j}+\frac{\partial}{\partial{Q_{j}}}\Big)$ and $\hat{a}_{j}^{\dagger}=\frac{1}{\sqrt{2}}\Big(Q_{j}-\frac{\partial}{\partial{Q_{j}}}\Big)$ for the annihilation and creation operators, respectively. 
The vector potential operator for the different radiation modes is given by $\hat{\vec{A}}_{j}(t) = \vec{z}_{j}\sqrt{\frac{\pi c^{2}}{\omega_{j}V}} \Big[\hat{a}_{j} e^{-\I\omega_{j} t}+\hat{a}^{\dagger}_{j}e^{\I\omega_{j} t}\Big]$, where $\vec{z}_{j}$ is the polarization direction and $V$ a quantization volume. 

The initial state of the electromagnetic field is given as a product of a coherent state (fundamental mode) and vacuum states (harmonic modes up to a cutoff~$M$), $\ket{G_{0}} = \ket{\alpha_{\rmL}} \bigotimes_{j}\ket{0_{j}}$ where $j\in \{2,\cdots, M\}$. In the position representation, this state can be expressed as $G_{0}(\vec{Q}) = \braket{\vec{Q}|G_{0}} = C \cdot\exp[-\frac{1}{2}(Q_{\rmL}-e^{\I\theta_{0}}\sqrt{2N_{0}})^{2}]\cdot\prod_{j}e^{-Q_{j}^{2}/2}$, where $\vec{Q} = (Q_{\rmL},Q_{2},\cdots,Q_{M})$, $\ket{\vec{Q}} = \ket{Q_{\rmL}}\bigotimes_{j}\ket{Q_{j}}$, $\theta_{0}$ is the internal field-phase, and $N_{0}\gg{}1$ is the initial average photon number of the driving laser mode, i.e., $\alpha_{\rmL} = \E^{\I\theta_{0}}\sqrt{N_{0}}$. Without loss of generality, we choose a reference frame in the optical phase space such that $\theta_{0}=0$ for the initial coherent state describing the laser mode. In this case, the related amplitude of the classically-described vector potential is $A_{0}=\sqrt{\frac{2\pi c^{2}}{\wL V}2N_{0}}$. Further, we assume in the following that the interaction strength is weak and the interaction time of the intense laser field with the semiconductor is sufficiently short. Then the momentum quadrature of the laser mode can be neglected in the interaction Hamiltonian, and the intense laser field operator can be considered as a local operator $\vec{z}_{\rmL}\hat{A}_{\rmL}(t) \equiv \hat{\vec{A}}_{\rmL}(t) \approx \vec{z}_{\rmL} \sqrt{\frac{2\pi c^{2}}{\wL V}} \cos{(\wL t)}\QL$. This allows us to write the $z$-projection of the vector operator for the whole system $\hat{A}(t)$ in the following way:
\begin{align}\label{eq:VectorPotential}
\hat{A}(t) &= \hat{A}_{\rmL}(t) + \sum_{j\geq{2}}\hat{A}_{j}(t)   \\
&= \sqrt{\frac{2\pi c^{2}}{\wL V}} \cos{(\wL t)}\QL + \sum_{j} \sqrt{\frac{\pi c^{2}}{\omega_{j}V}} \Big[\hat{a}_{j} e^{-\I\omega_{j} t}+\hat{a}^{\dagger}_{j}e^{\I\omega_{j} t}\Big]. \notag
\end{align}
For the initial field $\ket{G_{0}}$, the norm of the operators from Eq.~\eqref{eq:VectorPotential} fulfills the inequality $||\hat{A}_{\rmL}||\gg||\hat{A}_{j}||$, i.e., the pump field dominates the harmonic modes. This allows us to linearize the Hamiltonian in the TDSE~\eqref{eq:TDSE},
\begin{equation}\label{eq:TDSElin}
\I \frac{\partial}{\partial t} \ket{G} =\left[\nE \Ec\bigg(\frac{e}{c} \hat{A}_{\rmL}(t) \bigg) + \nE \sum_{j} \frac{e}{c} \hat{A}_{j}(t)\,\cdot\,\frac{\partial \Ec}{\partial K}\Big|_{K=\frac{e}{c} \hat{A}_{\rmL}}\right] \ket{G}.
\end{equation}
This is the main equation we obtain in this letter to describe the interaction of light (one dominant laser mode with the field local operator $\hat{A}_{\rmL} \sim \cos{(\wL t)}\QL$ and an arbitrary number of emitted light-modes labeled by the index $j$) with the semiconductor intraband excitation with band dispersion $\Ec$ at a quantum mechanical level.

Since Eq.~\eqref{eq:TDSElin} is linear with respect to non-local operators (the momentum quadrature operators $\hat{P}_{j} = -\I \partial_{Q_{j}}$ encoded in the ladder operators of the harmonic modes), it can be solved analytically by using standard mathematical approaches. For simplicity, we represent the band structure of the conduction band by a cosine potential (as commonly done), $\Ec(K) = \Eg[1  - \cos(\pi K/\Kc)]$, where $\Kc$ is an inverse lattice constant and $\Eg$ is the conduction band half-width and corresponds to the strength of the orbital interactions within the semiconductor. To analyze Eq.~\eqref{eq:TDSElin}, it is convenient to use the real-valued Jacobi–Anger expansions with the Bessel functions of the first kind $J_{k}$:
%
%
%
%
\begin{align*}
\cos\Big[ x_{1}\cos(x_{2})\Big] &= J_{0}(x_{1}) + 2\sum^{\infty}_{n=1} (-1)^{n}\,J_{2n}(x_{1}) \cos\big(2n\, x_{2}\big), \\
\sin\Big[ x_{1}\cos(x_{2}) \Big] &= -2\sum^{\infty}_{n=1} (-1)^{n}\,J_{2n-1}(x_{1}) \cos\big( (2n-1) x_{2}\big).
\end{align*}

Using the fact that the commutator of the Hamiltonian of the system with itself at different times is merely proportional to a function depending on time and the quadrature operator $\QL$, $[\hat{H}(t),\hat{H}(t')] = c(t,t';\QL)$, all higher-order commutators of the Baker-Campbell-Hausdorff formula vanish when the unitary time-evolution operator is constructed. Thus, the final state is given by:
\begin{align}
\ket{G} = \E^{\I f(t;\QL)}\, \exp \bigg\{ \sum_{j} \alpha_{j}(t;\QL) a_{j}^{\dagger} - [\alpha_{j}(t;\QL)]^{\dagger} a_{j} \bigg\} \ket{G_{0}}
\label{eq:TimeEvol}
\end{align}
where, 
\begin{align*}
f(t;\QL) &= \nE \Eg \Big[J_{0}\big( \gamma_{\rmL} \QL \big) - 1 \Big] (t-t_0) \\
&\quad + 2 \nE \Eg \sum_{n=1}^{\infty} (-1)^{n} J_{2n}\big( \gamma_{\rmL} \QL \big) \int\limits_{t_0}^{t}d\tau \cos(2n\, \wL\tau) \notag \\
&\quad + 2\nE^{2} \Eg^{2} \sum_{j} \sum_{n=1}^{\infty} \sum_{m=1}^{\infty} (-1)^{n+m} \gamma_{j}^{2} \times \notag \\ 
&\quad \int\limits_{t_0}^{t}\!\!\! d\tau \! \int\limits_{t_0}^{t}\!\!\! d\tau'  \Big[ \cos{\big((2n-1)\wL \tau \big)}\cos{\big((2m-1)\wL \tau' \big)} \times \notag\\
&\qquad \sin{\big(\omega_{j} (\tau'-\tau) \big)} \Big] J_{2n-1}\big( \gamma_{\rmL} \QL \big) J_{2m-1}\big( \gamma_{\rmL} \QL \big) \notag
\end{align*}
with abbreviation $\gamma_{\rmL/j} = \sqrt{2\pi^{3} e^{2}/(\omega_{\rmL/j}V\Kc^{2})}$ and
\begin{align}
 \alpha_{j}(t;\QL) &= \I \sqrt{2} \nE \Eg \sum_{n=1}^{\infty} (-1)^{n} \gamma_{j} J_{2n-1}\big( \gamma_{\rmL} \QL \big) \times \notag \\
 &\qquad \int\limits_{t_0}^{t}d\tau\, \cos{\big((2n-1)\wL \tau \big)} \E^{\I \omega_{j}\tau}.
\end{align}
The first exponential function in Eq.~\eqref{eq:TimeEvol} only contributes to the modification of the state of the driving laser mode due to the interaction with the conduction band current. The second exponential function corresponds to the harmonic generation via displacement operators with peaks of odd orders of $\omega_{\rmL}$ following the time integral dependencies in the expression for ${\alpha}_{j}(t;\QL)$. The generation of these modes is intertwined with further non-Gaussian modifications of the pump field given by the nonlinear $\QL$-dependencies of the coherent displacements. For the following investigations, it is also useful to study the solution in Eq.~\eqref{eq:TimeEvol} in the position quadrature representation,
\begin{align}
 G = \E^{\I f(t;Q_{\rmL})}\, \prod_{j} \E^{-\frac{1}{2}\big([\alpha_{j}(t;Q_{\rmL})]^{2} + |\alpha_{j}(t;Q_{\rmL})|^{2}\big)}\E^{\sqrt{2}\alpha_{j}(t;Q_{\rmL})Q_{j}} G_{0}.
\label{eq:GPosition}
\end{align}
%


\section{Driving field Classical limit}

\begin{figure}[]
\centering   
\includegraphics[scale=0.32]{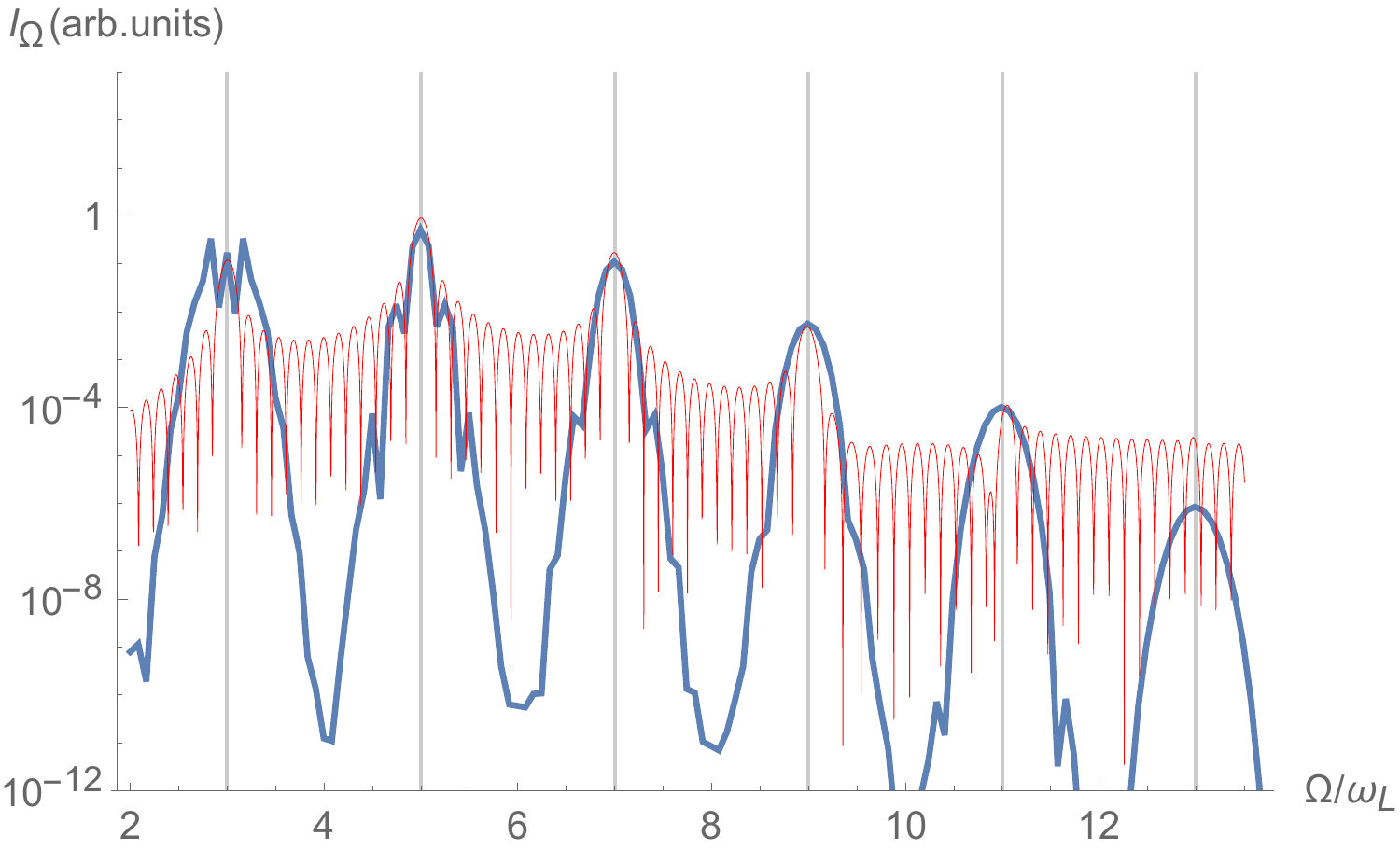} 
\caption{Intensity of intraband emission as a function of $\Omega/\wL$ for the given interaction time ($\sim{}3$ optical cycles at FWHM) and the given Bloch parameter $\wB/\wL=5.7$ obtained from our analytical solution (red) and the SMBE (blue). $\Omega$ is the frequency of the emitted radiation with clear picks around the higher-order harmonics given by $\Omega = j \wL$.}
\end{figure}

To discuss the properties and features of the solution of the TDSE, see Eq.~\eqref{eq:TimeEvol} or in position representation Eq.~\eqref{eq:GPosition}, we first consider the classical field limit for the intense non-perturbed driving laser radiation. Following the standard procedure, we calculate the quantum mechanical expectation values of the operators and neglect small quantum corrections. Since we actually have $\left\langle \QL \right\rangle_{G} = \left\langle \QL \right\rangle_{G_{0}} = \sqrt{2N_{0}}\gg{1}$ (see discussion in the next section), we have $\gamma_{\rmL}\QL \to \gamma_{\rmL}\braket{\QL} = \frac{\wB}{\wL}$ where $\wB = \frac{eA_{0}}{c\Kc}\wL$ is the Bloch frequency. Typical values of semiconductor crystals for the ratio $\wB/\wL$ are:
\begin{equation}\label{i8}
\frac{\wB}{\wL} = \frac{eA_{0}}{c\Kc} \sim \frac{\lambda \sqrt{I_{0}}}{\Kc},
\end{equation}
where $I_{0}$ is the laser field's peak intensity, $\lambda=\frac{2\pi{c}}{\wL}$ is the driving laser wavelength, and $\pi/\Kc$ is the lattice constant, typically in range of $4$-$6$ \AA. For instance, $\wB/\wL \approx 1$ for $I_{0}=5\cdot{}10^{11}$\,W/cm$^{2}$, $\lambda=1.44$ $\mu${m}, and $\pi/\Kc=4.4$ \AA. In previous experiments where high harmonic radiation was generated from various semiconductor targets \cite{s0,Yoshikawa2019} a typical ratio is in the range of up to 5 though it can be significantly higher when going to mid- or even far-infrared drivers \cite{Schubert2014}.

Using Eq.~\eqref{eq:GPosition}, with these parameters at hand, we calculate the harmonic amplitude ($m=2n-1$, and $m\geq{3}$): 
\begin{align}\label{i9}
 E_{\Omega} \Big( \frac{\wB}{\wL} \Big) &= \nE\Eg \,\Omega\sum_{m} J_{m} \Big( \frac{\wB}{\wL} \Big)\times \\
&\frac{\Omega\sin(\Omega\, t)\cos(m\, \wL\, t)-m \wL\cos(\Omega\, t)\sin(m\wL\, t)}{\Omega^{2}-m^{2}\wL^{2}}, \notag
\end{align}
where $t$ is the integration upper-limit, being the interaction time of the laser with the semiconductor and $\Omega$ is the frequency of the emitted radiation. 

The comparison of the emission intensity spectrum obtained from Eq.(\ref{i9}) and from the SMBE numerical calculations (conducted TDSE numerical calculations give a similar picture) is presented in Fig.1. The behavior of the odd-harmonic peak intensities coincides very well, however the background level (or lowest even-harmonic intensities) and harmonic widths differ. These differences can occur because the simplified analytical model does not include a time- and bloch vector-dependent conduction band population $\nE$ and assumes a monochromatic driving laser field with finite interaction duration, i.e., finite integration time.

\begin{figure}[]
	\centering   
	\includegraphics[scale=0.3]{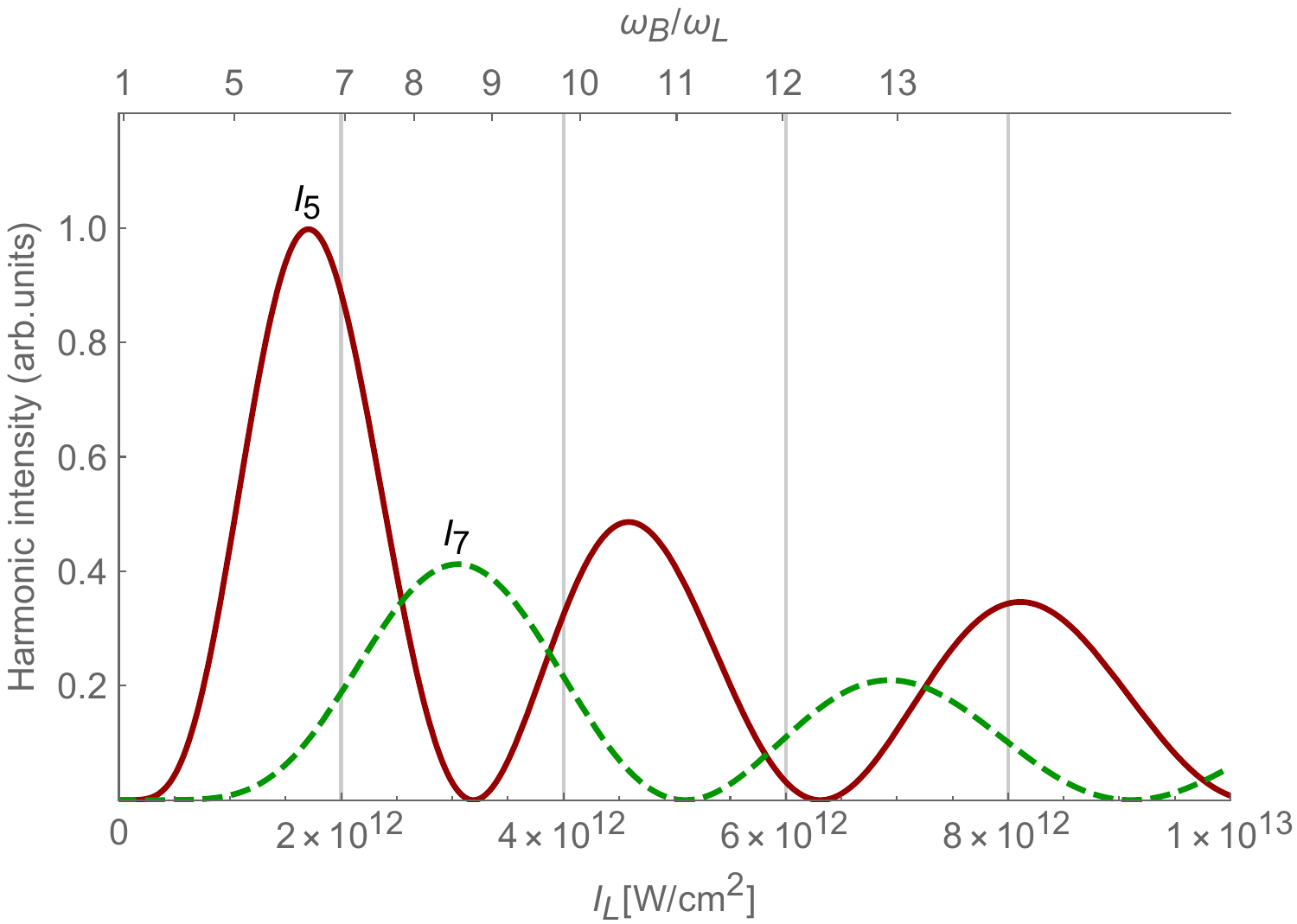} 
	\caption{Intensity of the 5-th and 7-th harmonic as a function of the laser intensity for $\lambda = 5$ $\mu${m} and $\pi/\Kc=4.4$ \AA.}
\label{fig:5-th}
\end{figure}

However, the good agreement of our theoretical approach with the numerical calculations for the odd-harmonic intensity dependence on the laser intensity can be utilized in various related applications. For instance, we calculate the dependence of the intensity of a certain intraband harmonic as a function of the laser intensity, using Eq.(\ref{i9}). As an example, we plot the dependence of the intensity of the 5-th and 7-th harmonic on the driving field's intensity in Fig.~\ref{fig:5-th}. We can identify several regions related to the different regimes of the generation (discussed for the 5-th harmonic in the following): 1) perturbative regime, when $\;I_{0}\lesssim{}10^{11}$ W/cm$^{2}$; 2) Linear growth with the fastest rate around $\;I_{0}\sim{}1\cdot{}10^{12}$ W/cm$^{2}$; and 3) the first local maximum at $\;I_{0}\sim{}1.5\cdot{}10^{12}$ W/cm$^{2}$. 

\section{Non-classical properties of emitted radiation}
For an explorative discussion concerning the non-classical features of light at the output of the laser-driven semiconductor, we focus on two points: 1) Non-coherent modification of the transmitted IR laser 
and 2) entanglement between the different modes, i.e., when the state cannot be written as a product state: $G(\vec{Q})\neq G_{\rmL}(Q_{\rmL}) \prod_{j}G_{j}(Q_{j})$, see also \cite{Ent}.

Our solution given in Eq.~\eqref{eq:TimeEvol} has a few remarkable properties. In general, the intricate unitary transformation, describing the evolution from the initial state $\ket{G_0} = \ket{\alpha_{\rmL}} \bigotimes_{j}\ket{0_{j}}$ to the final state $\ket{G}$ after the interaction generates non-Gaussian modifications of the fundamental mode according to the non-quadratic dependency on the quadrature operator $\QL$. This is in clear contrast to HHG in atoms. In the latter case, a depletion of the pump mode is observed but it remains a coherent state due to the linear interaction between the radiation field and the atoms see, e.g., the detailed analysis in \cite{Stammer, Lewenstein}. This can be traced back to the fact that the interaction Hamiltonian depends linearly on the field operator resulting in a multimode displacement operator. In our case, the nonlinear band dispersion allows for more involved properties due to nonlinear interactions between the modes. For instance, the unitary operator in Eq.~\eqref{eq:TimeEvol} is not described by a simple displacement but has the structure of a $\QL$-dependent displacement operator for the harmonics. Additionally, the pump field is altered according to a nontrivial $\QL$-dependent phase term. To investigate the properties of our solution, it is convenient to perform the analysis in the position representation, cf. Eq.~\eqref{eq:GPosition}.

Based on the assumption that the driving laser field is strong and negligibly perturbed, we directly obtain from Eq.~\eqref{eq:GPosition} that the average of $\QL$ remains unchanged, $\braket{\QL}_{G} = \sqrt{2N_{0}}$, within our approximations. In particular, this holds for the expectation value of any function of the position quadrature operator $\QL$, $\braket{f(\QL)}_{G} = \braket{f(\QL)}_{G_0}$. Nevertheless, other expectation values might be altered but deviations from the coherent state scale sufficiently small  for weak interactions of the pump laser with the semiconductor. We provide an explicit example for the momentum quadrature below. 

To analyze the quantum properties of the final state $\ket{G}$, we expand the Bessel functions $J_{k}(\alpha)$ in Eq.~\eqref{eq:GPosition}. Based on the regimes considered in the previous section in Fig.2, we can identify various situations depending on the ratio $\frac{\omega_{B}}{\wL}$. For instance, for $\frac{\omega_{B}}{\wL}\ll{1}$, we are in the well known perturbative regime when $I_{\Omega}\sim{}I_{\wL}^{2n+1}$, since $J_{2n+1}(\gamma_{\rmL}Q_{\rmL})\sim{}(\gamma_{\rmL}Q_{\rmL})^{2n+1}$ for $\gamma_{\rmL}Q_{\rmL} = \frac{\omega_{B}}{\sqrt{2N_0}\wL}Q_{\rmL} \ll{1}$. In this case, the non-Gaussian modifications of the transmitted IR laser due to the excited intraband current is very small, the field state is almost unaffected and remains almost a coherent state.

In case we consider a second region with the fastest linear growth according to Fig.2, we can linearize the solution Eq.~\eqref{eq:GPosition} around $Q_{\rmL} \approx \sqrt{2N_{0}}$ and obtain the following linear approximation of the final state: $G\sim {G}_{0}(\vec{Q})e^{\delta_{3}Q_{\rmL}Q_{3}}\,e^{\delta_{5}Q_{\rmL}Q_{5}}\,...$\;, with constants $\delta_{j}$ depending on the parameters of the system, e.g., the band dispersion or the pump intensity. We also obtain this structure in case we assume that the relevant physics is dominated by the electrons in the middle of the Brillouin zone around the $\Gamma$ point, $\Ec(K) = \frac{\Eg K^{2}}{2\Kc^{2}} + \mathcal{O}(K^{4})$. In this regime, the complex parameters characterizing the displacement of the generated higher harmonics are directly proportional to the position quadrature of the pump field $\alpha_{j} \sim \gamma_{j} Q_{\rmL}$, i.e., the mean photon number of the harmonics scales proportional to the mean photon number $N_{0}$ of the pump laser. Of course, HHG is an inherent nonlinear process inducing a nonlinear relation between the mean photon number of the higher harmonics and the intensity of the pump mode stored in the parameter $\alpha_{\rmL}$, $\braket{\hat{N}_{j}} \sim \int \mathrm{d}Q_{\rmL} |\alpha_{j}(t;Q_{\rmL})|^{2}\E^{-(Q_{\rmL}-\sqrt{2}\alpha_{\rmL})^{2}}$. However, the nonlinear properties become subleading in this particular regime which can also be inferred from Fig.~\ref{fig:5-th}. Additionally, we obtain slight deviations for the exception value of the momentum quadrature $\hat{P}_{\rmL}$ compared to the initial state $\ket{G_{0}}$. However, these deviations are sufficiently small in accordance to our approximations. In particular, we have $\braket{\hat{P}_{\rmL}}_{G} = -\frac{1}{2} \nE \Eg \frac{\wB^{2}}{\sqrt{2N_0}\wL^{2}}t$ at lowest order in the interaction time $t$, the energy of the intraband current $\sim\nE\Eg$, and the dimensionless Bloch parameter $\frac{\wB}{\wL}$.



In order to describe the potential entanglement between the different frequency modes, we have to perform a conditioning on the generation of harmonic modes. Such a conditioning can be performed by the projection operator $\hat{\bm{1}} - \ket{G_{0}}\bra{G_{0}}$ \cite{Stammer}. Its orthogonal complement $\ket{G_{0}}\bra{G_{0}}$ projects on the subspace where no harmonic radiation has been produced via the process. Thus, in case higher harmonic modes are excited, which is trivially heralded by the emission of harmonic radiation, the correspondingly projected state is given by
\begin{align}
 \ket{G_{\mathrm{HHG}}} &= \big( \hat{\bm{1}} - \ket{G_{0}}\bra{G_{0}} \big) \ket{G} = \ket{G} - \braket{G_{0}|G}\, \ket{G_{0}} \\
 &= \ket{G} - \big\langle\alpha_{\rmL}\big| \E^{\I f(t,\QL) -\frac{1}{2} \sum_{j} \alpha_{j}^{\dagger}(t,\QL)\alpha_{j}(t,\QL)} \big| \alpha_{\rmL} \big\rangle \ket{G_{0}} \notag
\end{align}
The fact that this state is massively entangled between all optical modes present after HHG can again be conveniently explored in the position representation
\begin{align} \label{eq:G-HHG}
 \braket{\vec{Q}|G_{\mathrm{HHG}}} 
 &= \bra{\vec{Q}}\, \E^{\I f(t;Q_{\rmL})} \ket{\alpha_{\rmL}} \otimes_{j} \ket{\alpha_{j}(t;Q_{\rmL})}  \\
 &\quad - \bra{\vec{Q}} \int \mathrm{d}Q_{\rmL}\, \E^{\I f(t;Q_{\rmL})} |\braket{Q_{\rmL}|\alpha_{\rmL}}|^{2} \ket{\alpha_{\rmL}} \otimes_{j} \ket{0_{j}}.  \notag 
\end{align}
Equation~\eqref{eq:G-HHG} clearly demonstrates that the state conditioned on HHG cannot be written as a simple product of wave functions and is nonseparable. Therefore, the generation of higher harmonics via the interaction of the pump field with the nonlinear intraband structure of a semiconductor naturally produces an entanglement between all present field modes. In our specific case, the entangled state $\braket{\vec{Q}|G_{\mathrm{HHG}}}$ is on structural grounds similar to the entangled state produced in high harmonic generation in atoms \cite{Stammer2} or recently in solids within a Wannier-Bloch picture \cite{Rivera-Dean:2022tlk}. Apart from the different modification of the pump field, the higher order harmonics allow for a multipartite entangled coherent state. Such states provide a powerful resource for general quantum information processing \cite{Jeong}, e.g., for improved phase estimation \cite{Joo}, violations of Bell inequalities for continuous variables \cite{Gilchrist}, and quantum state engineering \cite{Walmsley}. The main advantage of the HHG process is that they can be produced over a broad spectral range from infrared to ultraviolet scales. The complex numbers $\alpha_{j}$ characterizing the displacements generally depend in a nonlinear way on the position quadrature of the fundamental mode allowing for large photon numbers in the harmonic modes. Already in the regime dominated by the minimum of the intraband potential, we obtain $|\alpha_{j}|^{2} \sim N_{0}$.

\section{Conclusions}
We have investigated the problem of multi-mode-light induced intraband excitations within a full quantum description. Within a first-order approximation, see Eq.~\eqref{eq:TDSElin}, we analyzed classical and nonclassical properties of the emitted radiation. In particular, HHG leads to entanglement between all optical modes in a natural way. On structural grounds, this is similar to HHG in atoms. A crucial difference arises due to the stronger dependence of the harmonics on the properties of the fundamental mode due to nonlinear interactions induced via the band dispersion. We carefully checked that our solution is consistent within the used approximations but emphasize that extended analyses, e.g., including nonlocal terms for the fundamental mode, might be valuable for deeper insights into the non-Gaussian modifications of the pump field. This might open new possibilities for the engineering of complex photonic resource states for quantum information processing.



\section{ACKNOWLEDGMENTS}
IG, CH, DK, UP, and SG acknowledge funding by Q-Hub Thüringen and the German Science Foundation DFG via SFB 1375 NOA (project number 398816777), projects A1 and C4. SG and IG also acknowledge funding from the ERC Consolidator grant QUEM-CHEM project number 772676. RS acknowledges funding by the European Union's Horizon 2020 Research and Innovation Action under Grant Agreement No. 899824 (SURQUID) and valuable discussions with Gregor Sauer and Fabian Steinlechner.


\end{document}